\documentclass[journal,12pt,onecolumn]{IEEEtran}
\usepackage[T1]{fontenc}
\usepackage{amsmath}
\usepackage{epsfig,amssymb,amsbsy,verbatim,subfigure,array}
\usepackage{pstricks,theorem,caption,url}
\usepackage{multicol,afterpage,mathtools,trfsigns} 
\usepackage{float}
\usepackage{multirow}
\usepackage{setspace}
\usepackage[margin=1in]{geometry}
\usepackage[compress]{cite}

\begin{document}
\title{Anywhere Decoding: Low-Overhead Uplink Interference Management for Wireless~Networks}
\author{Hamed Pezeshki, Masoumeh Sadeghi, Martin Haenggi, \textit{Fellow, IEEE}, \\and J. Nicholas Laneman, \textit{Fellow, IEEE}}

\maketitle

\begin{abstract}
Inter-cell interference (ICI) is one of the major performance-limiting factors in the context of modern cellular systems. To tackle ICI, coordinated multi-point (CoMP) schemes have been proposed as a key technology for next-generation mobile communication systems. 
Although CoMP schemes offer promising theoretical gains, their performance could degrade significantly because of practical issues such as limited backhaul. To address this issue, we explore a novel uplink interference management scheme called anywhere decoding, which requires exchanging just a few bits of information per coding interval among the base stations (BSs). In spite of the low overhead of anywhere decoding, we observe considerable gains in the outage probability performance of cell-edge users, compared to no cooperation between BSs. Additionally, asymptotic results of the outage probability for high-SNR regimes demonstrate that anywhere decoding schemes achieve full spatial diversity through multiple decoding opportunities, and they are within 1.5 dB of full cooperation.
\end{abstract}
%

\section{Introduction}
With the exponential growth in the demand for mobile data, wireless systems in general are experiencing densification of the wireless network elements that provide mobile data access. A notable example of wireless systems that have followed this densification trend is cellular systems, in which the high demand for data has been addressed through the introduction of heterogeneous cellular networks (HCNs) \cite{ghosh2012heterogeneous}. HCNs are a paradigm shift in the deployment of cellular network infrastructure, moving away from expensive high-power macro base stations mounted on towers to less expensive lower-power small cells mounted on buildings and light poles. Small cells include microcells, picocells, femtocells as well as distributed antenna systems, all of which are distinguished by their transmit power, coverage areas, physical size, backhaul, and propagation characteristics. Macrocells are typically interconnected through high-speed fiber optics links, whereas small cells are backhaul-constrained due to deployment limitations, putting constraints on the cooperation mechanisms.

As wireless networks become more and more dense, we expect higher-quality signal reception, thanks to reduced distance between the transmitters and desired receivers. However, due to scarcity of the spectrum, wireless systems have to reuse the available spectrum, which in turn leads to excessive interference. In the context of small cells, inter-cell interference (ICI) is one of the major performance-limiting factors, which has fueled research to develop interference management mechanisms and technologies. On the cellular standardization front, some of these interference management schemes have been unified into coordinated multipoint (CoMP) techniques \cite{irmer2011coordinated,nigam2014coordinated}, which is one of the key features of LTE-Advanced. It has been shown that dynamically coordinating the transmission and reception of signals across multiple cells could lead to significant gains in coverage and capacity by avoiding or mitigating interference \cite{marsch2011coordinated}. At a high level, these interference management mechanisms can be categorized as uplink (CoMP reception) and downlink (CoMP transmission) schemes. In this paper, we focus on uplink interference management schemes, motivated by the surge in the uplink traffic in the recent years, due to proliferation of smartphones and applications with user-generated content. 
\vspace{-2mm}
\subsection{Related Work}\label{relwork}
In this subsection, we review some of the major uplink interference management schemes in the literature. Uplink CoMP reception schemes can often be used with legacy terminals and are usually based on proprietary signal processing concepts, hence requiring little or no changes to standards. At a high-level, uplink interference mitigation schemes for cellular networks can be categorized into three major classes:

\textbf{Interference-aware detection}:
In this scheme, \emph{no cooperation} is necessary between the BSs. Instead, BSs estimate the channels of interfering terminals and either perform successive interference cancellation (SIC), take spatial characteristics of  interference into account in adjusting receive filters, i.e., interference rejection combining (IRC), or implement a combination of these two schemes (IRC+SIC) \cite{Moon2017}. 

\textbf{Joint multicell scheduling, interference prediction, or multicell link adaptation}:
In these \emph{cooperative} schemes, the BSs exchange information in order to coordinate resource usage and transmission strategies. Joint scheduling schemes belong to the broader class of so-called \emph{interference coordination} techniques, a notable example of which is inter-cell interference coordination (ICIC) \cite{hamza2013survey,zhang2014stochastic} in LTE and the enhanced ICIC in LTE-Advanced. Joint scheduling schemes generally require a relatively high backhaul load since multicell channel state information (CSI) of all cooperating BSs must be sent to a central scheduling unit. On the other hand, interference prediction or multicell link adaptation also leads to performance improvements in the uplink, at the expense of having low-latency backhaul links as a crucial pre-requisite \cite{marsch2011coordinated}. Therefore, generally, these multicell scheduling and link adaptation schemes require the exchange of channel information and/or scheduling decisions over the logical interfaces between~BSs, e.g., X2 in LTE.\\
\textbf{Joint multicell signal processing}:
For the \emph{cooperative} schemes within this category, there are  different centralized or decentralized decoding structures as well as different types of pre-processed received signals exchanged among the base stations. As two examples, distributed interference subtraction (DIS) exchanges the \emph{decoded messages} of the terminals over the backhaul links~\cite{balachandran2011nice}, and distributed antenna systems (DAS)~\cite{saleh1987distributed} exchange \emph{quantized receive signals} to enable centralized decoding, imposing a heavier load on backhaul, but at the same time providing a higher gain compared to DIS. 
\subsection{Contributions of this Work}
The backhaul has consistently been one of the major bottlenecks in the deployment of small cells. Hence, a common goal of the proposed multicell processing techniques for small cells is to optimize system performance with minimal information exchange between the BSs. Generally, one of the major hurdles for multicell processing techniques is the amount of overhead required for these schemes, which increases with the number of cooperating BSs. The reason for this increase is that including more BSs into a BS cluster requires more overhead for estimating CSI of the BSs/user euipments (UEs) in the cluster, which decreases the ratio of the transmitted data in a packet. For instance, it was shown in \cite{lee2015spectral} that a BS cluster with more than two BSs decreased the ergodic spectral efficiency when considering signaling overhead. Similar to \cite{park2016cooperative,baccelli2015stochastic} in which pairwise collaborative systems have been considered, we will also assume a pairwise collaborative system in this paper.

Setting aside interference coordination schemes (which are generally aimed at \emph{avoiding} interference preemptively, rather than \emph{mitigating} it), there are generally three classes of interference management schemes with different levels of data exchange among the BSs: 1) no data exchange, 2) exchange of decoded messages, and 3) exchange of quantized receive signals. In this work, we explore a scheme called \emph{anywhere decoding} \cite{pezeshki2015anywhere,phdthesis} that lies between classes 1 and 2 in terms of backhaul load, requiring \emph{just a few bits} per coding interval to be exchanged among the BSs. 
Unlike conventional association schemes in which the UEs are required to be decoded at pre-assigned BSs, anywhere decoding is based on the idea that for uplink transmissions it is not important at which BS the signal from a specific UE is decoded. Leveraging this concept allows us to have flexible decoding assignments through which BSs decode the UEs collaboratively. The BSs exchange indications of the decodability of the UEs using a few bits to help each other update the decoding assignments. We demonstrate considerable performance gains, specifically for UEs located at cell edges, where CoMP schemes are primarily intended. The asymptotic behavior of the outage probability in the high-SNR regime demonstrates that there is just a 1.5~dB gap between the performance of anywhere decoding and  full BS cooperation in which the BSs are connected through infinite-capacity, error-free backhaul links. 

In \cite{pezeshki2015anywhere}, we introduced anywhere decoding in the context of \emph{interference channels}, considering the \emph{capacity region} and \emph{common outage probability} as the key performance metrics, and using \emph{joint decoding} at the decoders. In this paper, we explore how anywhere decoding can be incorporated into \emph{practical cellular systems}. To this end, we introduce an anywhere decoding scheme that utilizes SIC at the BSs, rather than joint decoding as in \cite{pezeshki2015anywhere}.
\vspace{-2mm}
\subsection{Outline}
In Section \ref{sysmetric}, we summarize the system model and metrics used throughout the paper. In Section \ref{noi}, we perform an outage analysis for some uplink interference management schemes in the context of cellular systems. The schemes that we analyze are based on successive interference cancellation at the BSs. Later, in Section \ref{awdecoding}, we introduce, evaluate, and compare the performance of anywhere decoding with the schemes analyzed in Section \ref{noi}. We also discuss how the performance of anywhere decoding could improve in conjunction with the DIS scheme. Finally, in Section \ref{ppp}, we discuss how the performance of anywhere decoding is impacted by interference from outside the cooperating cells, using tools from stochastic geometry \cite{haenggi2009stochastic}.  
\section{System Model and Metrics}\label{sysmetric}
This section describes the cellular model and introduces the key performance metric as well as some notations that will  be used throughout the paper.
\subsection{System Model} \label{sysmod}
\begin{figure}[t]
	\centering
	\includegraphics[width=4.5in]{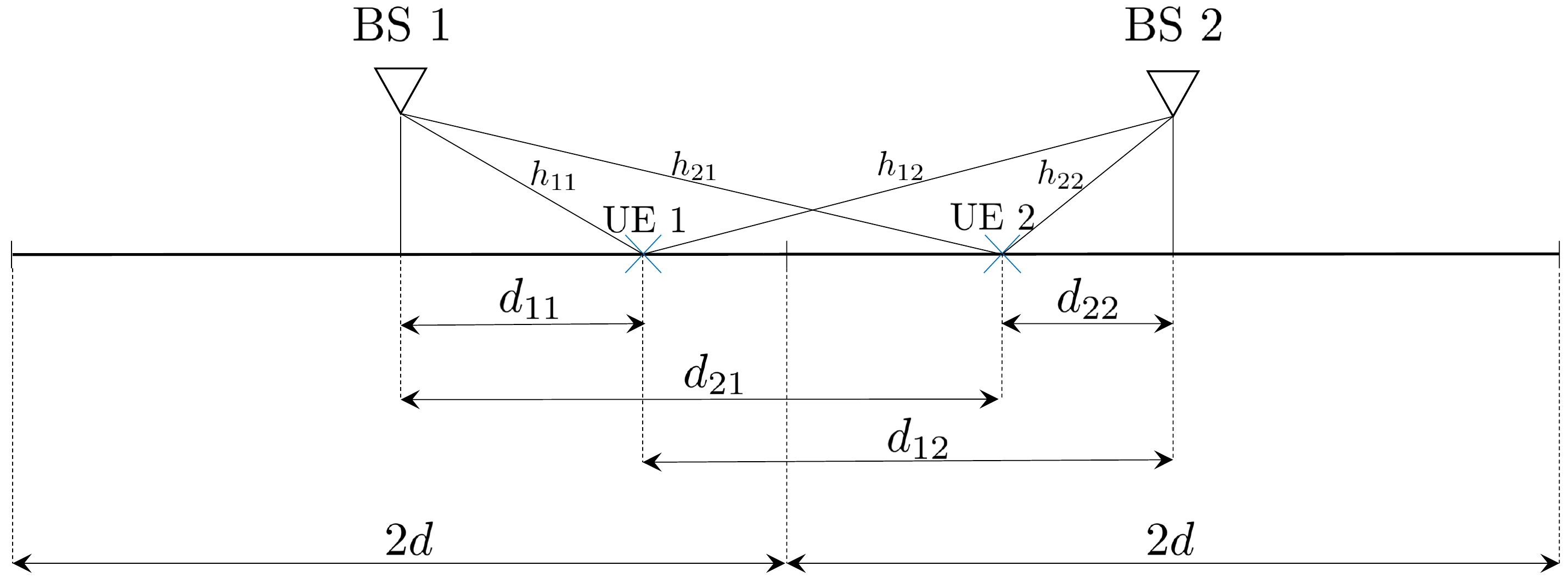}\\
	\caption{The 1-D grid-based cellular model.
	}\label{cellular}
\end{figure}

We consider a one-dimensional (1-D) cellular model, as depicted in Fig. \ref{cellular}, in which two cells are located  on a line, each including a BS and a single UE, and covering an interval of length $2d$. The left and right cells and their corresponding BS and UE are indexed by 1 and 2, respectively. We refer to cells 1 and 2 as \emph{cooperating cells} in the remainder of the paper. We consider frame synchronous uplink transmissions, incorporate path loss and Rayleigh fading in our model, and we neglect lognormal shadowing for simplicity. The channel gain between the $i$-th UE and the $j$-th BS is denoted by $h_{ij}$, and $h_{ij}=\frac{g_{ij}}{\sqrt{1+d_{ij}^\alpha}}$, where $g_{ij}$ is zero-mean circularly symmetric complex Gaussian with unit variance, $\alpha$ is the path loss exponent, and $d_{ij}$ is the ground distance from UE~$i$ to BS~$j$. The channel gains $h_{ij}$, $i,j=1,2$ are independent, but \emph{not} identically distributed due to path loss. Let us define $\lambda_{ij}\triangleq 1+d_{ij}^\alpha$, so that $\left|h_{ij}\right|^2\sim \exp(\lambda_{ij})$, i.e., $\left|h_{ij}\right|^2$ is exponentially distributed with mean $1/\lambda_{ij}$.

We denote the displacement of UE~1 from BS~1, and of UE~2 from BS~2, by $z$ and $t$, respectively. Considering the BS locations as the respective origin, and assuming that the positive direction is from left to right, we have  $-d\leq z,t\leq d$. Now, we can write the distances in terms of $z$ and $t$: $d_{11}=|z|$, $d_{22}=|t|$, $d_{12}=2d-z$, and $d_{21}=2d+t$. We assume that UE~1 and UE~2 transmit with rates $R_1$ and $R_2$, respectively. In the next sections, we will consider different scenarios regarding the respective locations of the users as well as their transmit powers.
\subsection{Metric and Notations}
As opposed to \cite{pezeshki2015anywhere}, where we considered common metrics such as the common outage probability, in this paper, we consider the \emph{individual outage probability} as the key metric. Let us consider a point-to-point transmission in which UE~$i$ transmits to BS~$j$ with transmit power $P_i$ and transmission rate $R_i$, and define $\theta_i=2^{R_i}-1$. By incorporating the impact of fading and Gaussian noise, the outage probability of UE~$i$ at BS~$j$ is defined as \cite{ozarow1994information}
\begin{equation}
\mathbb{P}\left(\log_2 \left(1+\frac{P_i|h_{ij}|^2}{N_0}\right)<R_i\right)= \mathbb{P}\left(\frac{P_i|h_{ij}|^2}{N_0}<\theta_i\right),\quad i,j\in\{1,2\},
\end{equation}
where $N_0$ is the noise power. The quantity $P_i|h_{ij}|^2/N_0$ is the received signal-to-noise ratio (SNR). In the remainder of the paper, we normalize $N_0$ to one, without loss of generality. Now, consider the case of concurrent transmissions from UE~$i$ and UE~$i'$, and BS~$j$ intends to decode UE~$i$ treating the interference from UE~$i'$ as noise. In this case, the outage probability is
\begin{equation}
 \mathbb{P}\left(\frac{P_i|h_{ij}|^2}{1+P_{i'}|h_{i'j}|^2}<\theta_i\right),\quad i,j\in\{1,2\},~,i'=3-i,
\end{equation}
where the quantity $P_i|h_{ij}|^2/(1+P_j|h_{i'j}|^2)$ is the received signal-to-interference-plus-noise ratio (SINR), and $\theta_i$ is called the SINR decoding threshold. In order for the transmission of UE~$i$ to be decodable at BS~$j$, the SINR of UE~$i$ at BS~$j$ needs to exceed the SINR decoding threshold. The asymptotic case of $\theta_i\rightarrow 0$, which will be considered in the remainder of the paper, refers to the high-reliability regime. 

For compactness in the remainder of the paper, we define the following notation:
\begin{itemize}
	\item $\mathcal{E}_{ij}$: The event that UE~$i$ is decoded successfully at BS~$j$, treating the other user, UE~$i'$ as noise, i.e.,
	\begin{equation}
	\mathcal{E}_{ij}: \frac{P_i|h_{ij}|^2}{1+P_{i'}|h_{i'j}|^2}\geq\theta_i,\quad i,j\in\{1,2\}, ~i'=3-i.
	\end{equation}
	The two events $\mathcal{E}_{ij}$ and $\mathcal{E}_{ij'}$ are independent if $j\neq j'$.
	\item $\mathcal{A}_{ij}$: The event that UE~$i$ is decoded successfully at BS~$j$, treating UE~$i'$ as noise \emph{or} using successive interference cancellation, i.e., 
	\begin{equation}
	\mathcal{A}_{ij}=E_{ij}\cup \left(E_{i'j}\cap \{P_i|h_{ij}|^2\geq\theta_i\}\right).
	\end{equation}
	\item The complement of event $\mathcal{E}$ is denoted by $\mathcal{E}^c$.
	\item The intersection of the two events $\mathcal{A}$ and $\mathcal{B}$ is sometimes denoted by $\mathcal{A}\mathcal{B}$.
\end{itemize}
\section{Outage Analysis of Some Interference Mitigation Schemes}\label{noi}
This section considers several uplink interference mitigation schemes in the literature and characterizes their outage probabilities. To this end, we begin with two assumptions: fixed user locations and no power control at the UEs. Later in this section, we relax these assumptions by randomizing the user locations and incorporating power control at the~UEs. 
\subsection{Fixed UE Locations, No Power Control}\label{npcfl}
We assume that the locations of the UEs are fixed and the UEs transmit with equal power, i.e., $P_1=P_2=P$. The goal of this subsection is to evaluate the performance of different decoding schemes based on these assumptions.
\subsubsection{Successive Interference Cancellation with Association Based on Maximum Average Received Power (MARP)}
We consider an association policy that assigns each UE to the BS at which it has the \emph{largest average receive power}. 
Equivalently in our model, the UEs are connected to their closest BSs, and we have a static long-term  association, where each BS is interested in decoding its corresponding signal of interest (SoI) coming from its associated UE. Here is how the MARP scheme works in the context of the 2-BS, 2-UE model introduced in Section \ref{sysmod}:

First, each BS attempts to decode its SoI, treating the interferer as noise. If decoding is successful, the process concludes. If decoding is not successful, the BS attempts to decode the interferer, treating the SoI as noise. If the BS can decode the interferer, it subtracts off the interferer from the received signal, and again tries to decode the SoI.

We denote the SINR at the BSs corresponding to the MARP association policy by $\text{SINR}^\text{MARP}$. In this case, if we use SIC at BS~1, there are two options for the uplink SINR of UE~1, given~by
\begin{equation}\label{sinrmarp}
\rm{SINR}_1^{\rm{MARP}} =
\left\{
\begin{array}{ll}
P|h_{11}|^2  & \mbox{if}\quad \mathcal{E}_{21} \\
\frac{P|h_{11}|^2}{1+P|h_{21}|^2} &~~ \quad \text{otherwise}.
\end{array}
\right.
\end{equation}
Equation \eqref{sinrmarp} suggests that if BS~1 can decode the interference coming from UE~2, it can cancel UE~2's signal from its overall received signal to improve UE~1's decodability. Otherwise, BS~1 will treat the signal coming from UE~2 as noise.
The outage event for UE~1 using this scheme, denoted by $\mathcal{O}_1^{\text{MARP}}$, can be written~as~follows:
\begin{equation}\nonumber
\mathcal{O}_1^{\text{MARP}}=\mathcal{A}_{11}^c=\mathcal{E}_{11}^c\cap \left(\mathcal{E}_{21}^c\cup \{P|h_{11}|^2<\theta_1\}\right).
\end{equation}
Now, the outage probability for UE~1 can be derived as follows
\begin{align}
P_{\rm{out}_1}^{\rm{MARP}}=\mathbb{P}\left(\mathcal{O}_1^{\text{MARP}}\right)&=\mathbb{P}(\mathcal{E}_{11}^c\mathcal{E}_{21}^c)+\mathbb{P}(\mathcal{E}_{11}^c\cap\{ P|h_{11}|^2<\theta_1\})-\mathbb{P}(\mathcal{E}_{11}^c\mathcal{E}_{21}^c\cap \{P|h_{11}|^2<\theta_1\})\nonumber\\
&=\mathbb{P}(\mathcal{E}_{11}^c\mathcal{E}_{21}^c)+\mathbb{P}(P|h_{11}|^2<\theta_1)-\mathbb{P}(\mathcal{E}_{21}^c\cap \{P|h_{11}|^2<\theta_1\})\label{marpout1}\\
&=\left\{
\begin{array}{ll}
f(\lambda_{11},\lambda_{21},\theta_2;x)  & \mbox{if}\quad x\geq1/\theta_2\\
g(\lambda_{11},\lambda_{21},\theta_2;x) &  \text{otherwise},
\end{array}\label{marpout2}
\right.
 \intertext{where}
f(a,b,c;x)={}&1-{\rm e}^{-\frac{ax}{P}}\left(\frac{b}{ax+b}+\frac{ae^{-\frac{bc}{P}(x+1)}}{a+bc}\right)\label{marp},\\
g(a,b,c;x)
={}&f(a,b,c;x)+\exp\left(\frac{bc(1+x)+ax(1+c)}{P(c x-1)}\right)\frac{ab(1-c x)}{(a+bc)(ax+b)},
\end{align}
where the second term in \eqref{marpout1} results from the fact that the event $P|h_{11}|^2<\theta_1$ is a subset of the event $\mathcal{E}_{11}^c$. 

Asymptotically as $\theta_1\rightarrow 0$, the outage probability for UE~1 behaves as
\begin{equation}\label{marpasymp1}
P_{\rm{out}_1}^{\rm{MARP}}(\theta_1)\overset{\theta_1\rightarrow 0}\sim\left(\frac{1}{\lambda_{21}}\left(1-{\rm e}^{-\frac{\lambda_{21}\theta_2}{P}}\right)+\frac{1}{P}\left(1-\theta_2{\rm e}^{-\frac{\lambda_{21}\theta_2}{P}}\right)\right)\lambda_{11}\theta_1,
\end{equation} 
where $\sim$ denotes asymptotic equality.

For the symmetric case in which the UEs are transmitting at the same rate, i.e., $R_1=R_2=R$, we have $\theta_1=\theta_2=\theta=2^R-1$, the outage probability is $P_{\rm{out}_1}^{\rm{MARP,~sym}}(\theta)$, which can be expressed~as
\begin{equation}\label{marpoutsym}
P_{\rm{out}_1}^{\rm{MARP,~sym}}(\theta)=
\left\{
\begin{array}{ll}
f(\lambda_{11},\lambda_{21},\theta;\theta)  & \mbox{if}\quad \theta\geq1\\
g(\lambda_{11},\lambda_{21},\theta;\theta) & ~~\quad \text{otherwise},
\end{array}
\right.
\end{equation}
and the asymptotic outage probability in this case behaves as
\begin{equation}\label{marpasymp2}
P_{\rm{out}_1}^{\rm{MARP}}(\theta)\overset{\theta\rightarrow 0}\sim\frac{\lambda_{11}}{P}\theta.
\end{equation}
We infer from \eqref{marpasymp1} that, if $\theta_1\rightarrow 0$, we observe the effect of the interference from UE~2 in the asymptotic regime; on the other hand, \eqref{marpasymp2} demonstrates that the interference will be canceled if $\theta_1=\theta_2=\theta\rightarrow 0$, thanks to $\theta_2$ approaching zero and the use of SIC.

 MARP is a non-cooperative scheme in which there is no data exchange between the BSs over backhaul links, and it lies under the umbrella of \emph{interference-aware detection} schemes discussed in Section \ref{relwork}. We will occasionally refer to MARP as the baseline scheme in the remainder of the paper.

\subsubsection{Distributed Interference Subtraction (DIS)}
In this subsection, we review a scheme that allows data exchange between the BSs, with MARP as the association policy. Looking back at Fig.~\ref{cellular},  assume that the signal from UE~1 is decodable at its associated BS, i.e., BS~1, but is not strong enough to be decoded at BS~2. Therefore, the de facto option that BS~2 will have in terms of decoding UE~2 is to treat the signal coming from UE~1 as noise. Through DIS, the decoded message of UE~1 at BS~1 can be sent to BS~2 over the backhaul link. BS~2 can reconstruct the received signal from UE~1, assuming that BS~2 knows the channel from UE~1 to itself, and subtract it from its overall received signal to improve the SINR of UE~2 at BS~2. Let us denote the SINR at the BSs corresponding to DIS by $\rm{SINR}^{\rm{DIS}}$. In this case, the SINR of UE~1 is
\begin{equation}\label{sinrdis}
\rm{SINR}_1^{\rm{DIS}} =
\left\{
\begin{array}{ll}
P|h_{11}|^2  & \mbox{if}\quad \mathcal{E}_{21} \cup \mathcal{E}_{22}\\
\frac{P|h_{11}|^2}{1+P|h_{21}|^2}  & ~~\quad \text{otherwise}.
\end{array}
\right.
\end{equation}
As we see in \eqref{sinrdis}, $|h_{12}|^2$ does not play a role in the SINR for UE~1, because even if BS~2 can decode UE~1, it does not report UE~1's message to the network.

It can be verified that the outage event for UE~1 using the DIS scheme can be written~as
\begin{equation}
\mathcal{O}_1^{\rm{DIS}}=\mathcal{A}_{11}^c\cap \left(\mathcal{A}_{22}^c\cup \{P|h_{11}|^2<\theta_1\}\right).
\end{equation}
Compared to MARP, it is clear that the outage probability for the DIS scheme is smaller, since the outage event for DIS is a subset of $\mathcal{A}_{11}^c$. Now, we compute the outage probability for UE~1 using the DIS scheme
\begin{align}
\mathbb{P}(\mathcal{O}_1^{\rm{DIS}})&=\mathbb{P}(\mathcal{A}_{11}^c\mathcal{A}_{22}^c)+\mathbb{P}(\mathcal{A}_{11}^c\cap \{P|h_{11}|^2<\theta_1\})-\mathbb{P}(\mathcal{A}_{11}^c\mathcal{A}_{22}^c\cap \{P|h_{11}|^2<\theta_1\})\nonumber\\
&=\mathbb{P}(\mathcal{A}_{11}^c)\mathbb{P}(\mathcal{A}_{22}^c)+\mathbb{P}(P|h_{11}|^2<\theta_1)-\mathbb{P}(P|h_{11}|^2<\theta_1)\mathbb{P}(\mathcal{A}_{22}^c),\label{disout2}
\end{align}
where the first term in \eqref{disout2} results from the fact that $\mathcal{A}_{11}$ and $\mathcal{A}_{22}$ are independent events, and it can be verified that the event $P|h_{11}|^2<\theta_1$ is a subset of the event $\mathcal{A}_{11}^c$, which results in the second and third terms in \eqref{disout2}. After calculating the probabilities of the events in \eqref{disout2}, we derive the outage probability for UE~1 as
\begin{align}
P_{\rm{out}_1}^{\rm{DIS}}(\theta_1)
&=\left\{
\begin{array}{ll}
f_{\rm{DIS}}(\theta_1)  & \mbox{if}\quad \theta_1\geq1/\theta_2\\
g_{\rm{DIS}}(\theta_1) & ~~\quad \text{otherwise},
\end{array}
\right.\\ \intertext{where}
f_{\rm{DIS}}(\theta_1)=f(\lambda_{11},\lambda_{21}&,\theta_2;\theta_1)f(\lambda_{22},\lambda_{12},\theta_2;\theta_1)+(1-{\rm e}^{-\frac{\lambda_{11}\theta_1}{P}})(1-f(\lambda_{22},\lambda_{12},\theta_2;\theta_1)),\nonumber\\
g_{\rm{DIS}}(\theta_1)=g(\lambda_{11},\lambda_{21}&,\theta_2;\theta_1)g(\lambda_{22},\lambda_{12},\theta_2;\theta_1)+(1-{\rm e}^{-\frac{\lambda_{11}\theta_1}{P}})(1-g(\lambda_{22},\lambda_{12},\theta_2;\theta_1)).\nonumber
\end{align}
\subsubsection{Association Based on Maximum Instantaneous SINR (MIS)}
In the MARP and DIS schemes, the fading random variables are averaged in the association policy. Considering fading, a UE could be associated with a BS that is not necessarily the closest to the UE, but instantaneously provides the highest UL SINR. For downlink transmissions some works have assumed that the UEs connect to the BSs offering the highest \emph{instantaneous} downlink SINR \cite{dhillon2012modeling}, mainly for deriving a bound on the performance of the system. However, considering the notion of decoupled uplink-downlink associations \cite{singh2015joint}, we evaluate the instantaneous UL SINR as a criterion for uplink association as~in \cite{wildemeersch2014successive}.

In this association policy, we have short-term BS-UE assignments based on instantaneous realizations of the channel gains. 
The MIS scheme outperforms the MARP scheme in terms of the UL SINR, at the expense of additional complexity and overhead. Let us denote the SINR at the BSs corresponding to the MIS association policy by $\text{SINR}^\text{MIS}$. If we utilize SIC at the BSs, the SINR of UE~1 will be given by

\begin{equation}\label{sinrmis2}
\text{SINR}_1^{\rm{MIS}} =
\left\{
\begin{array}{ll}
P|h_{11}|^2& \mbox{if}\quad \mathcal{E}_{21}\quad\text{and}\quad \frac{P|h_{11}|^2}{1+P|h_{21}|^2}>\frac{P|h_{12}|^2}{1+P|h_{22}|^2}\\
\frac{P|h_{11}|^2}{1+P|h_{21}|^2}  & \mbox{if}\quad \mathcal{E}_{21}^c\quad \text{and}\quad\frac{P|h_{11}|^2}{1+P|h_{21}|^2}>\frac{P|h_{12}|^2}{1+P|h_{22}|^2}\\
P|h_{12}|^2  & \mbox{if}\quad \mathcal{E}_{22}\quad\text{and}\quad \frac{P|h_{11}|^2}{1+P|h_{21}|^2}<\frac{P|h_{12}|^2}{1+P|h_{22}|^2}\\
\frac{P|h_{12}|^2}{1+P|h_{22}|^2} & \mbox{if}\quad \mathcal{E}_{22}^c\quad \text{and}\quad\frac{P|h_{11}|^2}{1+P|h_{21}|^2}<\frac{P|h_{12}|^2}{1+P|h_{22}|^2}.
\end{array}
\right.
\end{equation}

To understand the SINR expression in \eqref{sinrmis2}, if the instantaneous receive UL SINR of  UE~1 is larger at BS~1 than BS~2, i.e., $\frac{P|h_{11}|^2}{1+P|h_{21}|^2}>\frac{P|h_{12}|^2}{1+P|h_{22}|^2}$, then UE~1 will be associated to BS~1. Now, there are two options in terms of the decodability of UE~2 at BS~1. If UE~2 is decodable at BS~1, i.e., event $E_{21}$ occurs, $\text{SINR}_1^{\rm{MIS}}=P|h_{11}|^2$. Otherwise, BS~1 will have to treat the signal coming from UE~2 as noise, i.e., $\text{SINR}_1^{\rm{MIS}}=\frac{P|h_{11}|^2}{1+P|h_{21}|^2}$. A similar argument holds true if UE~1 is assigned to BS~2, which leads to the third or fourth expressions for $\text{SINR}_1^\text{MIS}$ in \eqref{sinrmis2}. 

Using the MIS scheme, the outage event for UE~1  can be written as
\begin{equation}\label{misout}
\mathcal{O}_1^{\rm{MIS}}=
\left\{
\begin{array}{ll}
\mathcal{A}_{11}^c& \mbox{if}\quad \frac{P|h_{11}|^2}{1+P|h_{21}|^2}>\frac{P|h_{12}|^2}{1+P|h_{22}|^2}\\
\mathcal{A}_{12}^c& ~~\quad \text{otherwise}.
\end{array}
\right.
\end{equation}
The outage probability of for this scheme will be explored numerically in Section \ref{comparison}.
\subsubsection{Minimum Mean Square Error-Successive Interference Cancellation (MMSE-SIC)}
We discuss a bound on the best achievable performance through an SIC-based scheme, called MMSE-SIC, assuming that the BSs are connected through infinite-capacity backhaul links. With these capabilities, the two BSs mimic a single two-antenna BS, i.e., an ideal DAS scheme. The SINR for UE~1 in this case is given by \cite{tse2005fundamentals}
\begin{equation}\label{sinrmmse}
\text{SINR}_1^\text{MMSE-SIC} =
\left\{
\begin{array}{ll}
P||\mathbf{h}_1||^2=P(|h_{11}|^2+|h_{12}|^2)  & \mbox{if}\quad P\mathbf{h}_2^*(\mathbf{I}+P\mathbf{h}_1\mathbf{h}_1^*)^{-1}\mathbf{h}_2>2^{R_2}-1\\
P\mathbf{h}_1^*(\mathbf{I}+P\mathbf{h}_2\mathbf{h}_2^*)^{-1}\mathbf{h}_1  & ~~\quad \text{otherwise},
\end{array}
\right.
\end{equation}
and
\begin{equation}\label{chvector}
\mathbf{h}_1= \begin{bmatrix} 
h_{11} \\
h_{12}
\end{bmatrix}, \quad \mathbf{h}_2=\begin{bmatrix} 
h_{21} \\
h_{22}
\end{bmatrix}.
\end{equation}
For the MMSE-SIC scheme, we omit the lengthy analysis, and we rely on numerical simulations in Section \ref{comparison}.

\subsection{Random UE Locations with Power Control} \label{rlpc}
In this subsection, we consider the system model discussed in Section \ref{sysmod}, with the following two additions:

\begin{itemize}
	\item We consider uplink power control (full path loss compensation) for the users within the two adjacent cells, i.e., cells 1 and 2. Specifically, if we denote the transmit power of a UE as $P_{\rm t}$, and the target received power at the associated BS as $P_{\rm r}$, we have $P_{\rm r}=A P_{\rm t}d_{\mathrm{UE-BS}}^{-\alpha}$, where $A$ is the propagation constant. We assume $A=1$ without loss of generality.
	\item The two users under consideration, i.e., UE~1 and UE~2 are located randomly in an interval, i.e., $Z\sim \mathcal{U}[d_1,d_1']$, and $T\sim \mathcal{U}[d_2,d_2']$, where $d_i<d_i'$ and $-d<d_i,d_i'<d, i\in\{1,2\}$.
	\end{itemize}
To incorporate power control, we can reuse the outage probability expressions derived in Section \ref{npcfl} considering a modified system model. For instance, instead of considering a controlled transmit power of $P\lambda_{11}$ at UE~1, which means its corresponding received power at BS~1 and BS~2 are $P$ and $P\lambda_{11}/\lambda_{12}$, respectively, we consider an equivalent system in which the transmit power is $P$, and the power control is incorporated into the path losses. Specifically, let us denote the updated values for $\lambda_{ij}$ by $\lambda_{ij}^P$, so that $\lambda_{11}^P=1$, $\lambda_{12}^P=\lambda_{12}/\lambda_{11}$, $\lambda_{21}^P=\lambda_{21}/\lambda_{22}$, and $\lambda_{22}^P=1$. 

To incorporate the effect of randomness in the UE locations, we can average the outage probability expressions derived in Section \ref{npcfl}. Let us denote the outage probabilities of UE~1 for fixed and random UE locations (for any of the decoding schemes) as $P_{\text{out}_1,f}$ and $P_{\text{out}_1,r}$, respectively. In this case, assuming that UE~1 and UE~2 are located in the intervals $[d_1,d_1']$ and $[d_2,d_2']$, respectively, we have

\begin{equation}
P_{\text{out}_1,r}=\int_{d_2}^{d_2'}\int_{d_1}^{d_1'}P_{\text{out}_1,f}(z,t)f_{ZT}(z,t){\rm d} z{\rm d} t,
\end{equation}
where $f_{ZT}(z,t)$ is the joint pdf of the locations of UE~1 and UE~2. 

\section{Uplink Interference Mitigation Using Anywhere Decoding}\label{awdecoding}
In this section, we describe how the anywhere decoding algorithm \cite{pezeshki2015anywhere} works using SIC at the BSs. As opposed to the non-cooperative scheme, in which each BS has only a single SoI, in the anywhere decoding scheme, there are two SoIs for the 2-BS, 2-UE model introduced in Section \ref{sysmod}: the primary signal of interest (PSoI), which is the UE with the higher long-term average received power, and the secondary signal of interest (SSoI), which is the UE with the lower long-term average received power. The signals from UE~1 and UE~2 are considered to be the PSoI and SSoI for BS~1, respectively. Throughout this section, we present the results for fixed UE locations and no power control at the UEs. We can extend the results to support random UE locations and power control at the UEs, by following the approach in Section \ref{rlpc}.
\vspace{-10mm}
\subsection{Anywhere Decoding  with Successive Interference Cancellation (AW+SIC)}
\begin{figure}[t]
	\centering
	\includegraphics[width=5.2in]{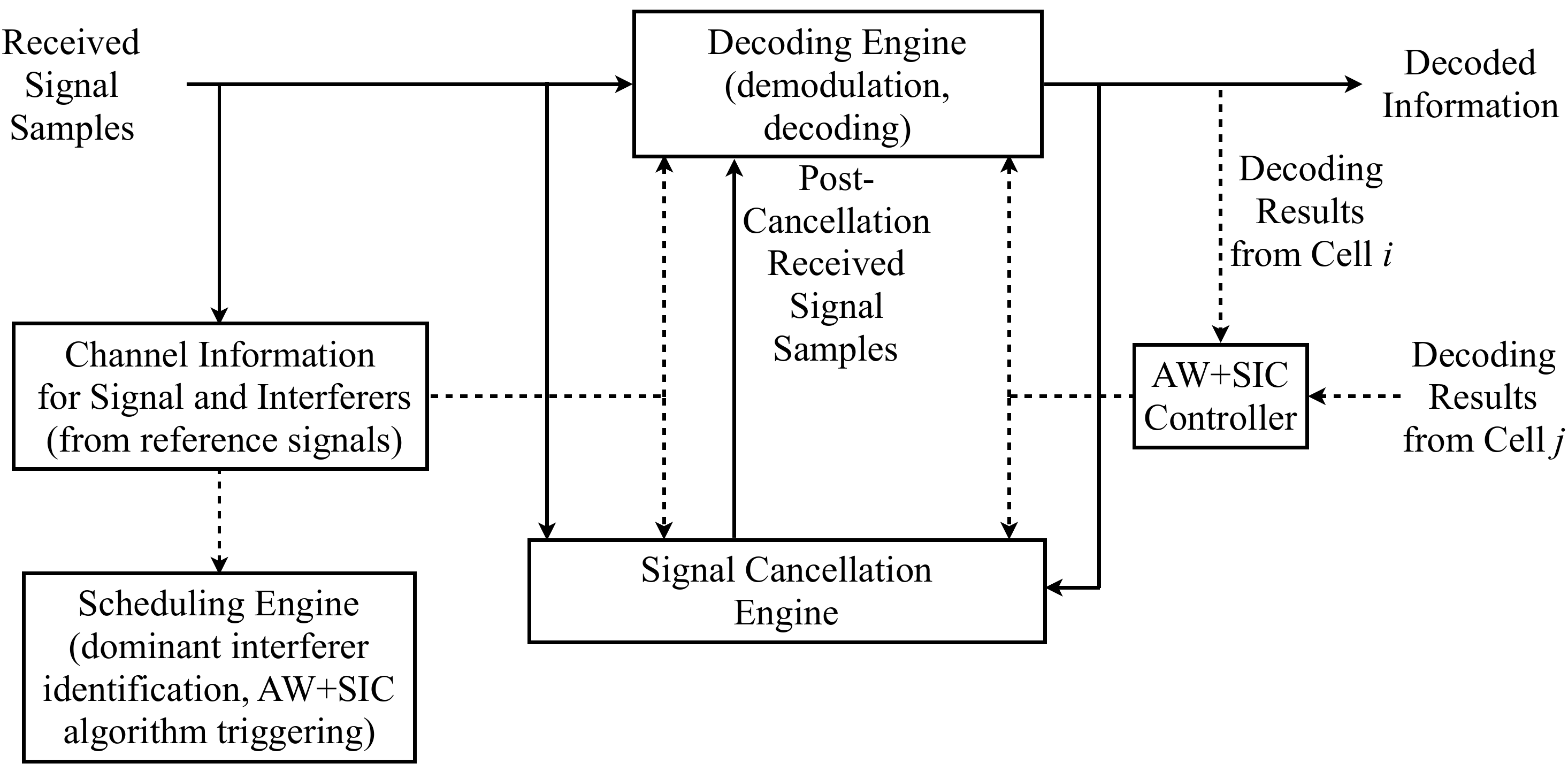}\\
	\caption{AW+SIC functional block diagram for BS~$i$. The AW+SIC controller determines the next decoding assignments based on the decoding results from the cooperating cells.
	}\label{AW_block_diagram}
\end{figure}

We explain the anywhere decoding algorithm in the context of our 2-BS, 2-UE model. Let us index the cooperating cells as $i$ and $j$. The three-step anywhere decoding algorithm has been summarized in Table \ref{awalgorithm}, where a controller has an initial decoding assignment for BS~$i$ and BS~$j$, and determines the future decoding assignments based on the decoding results of previous steps. This controller could be a separate entity, connected to BS~$i$ and BS~$j$, or it could be a part of each of the BSs.  To make the expressions concise, we have used the following notations:
\begin{itemize}
	\item BS~$i$: UE~$i$/UE~$j$ denotes a decoding assignment in which BS~$i$ decodes UE~$i$ treating UE~$j$ as noise.
	\item We represent the results of the decoding assignments by bits with ``1'' indicating that decoding is successful, and ``0'' indicating that decoding is not successful. Because we have two BSs, we can represent the decoding results by two bits, where the first and second bit denote the decoding result for BS~$i$ and BS~$j$, respectively. Since we have a decoding assignment for a single BS in the third step of the algorithm, we represent the decoding result by a single bit.
\end{itemize}
A functional block diagram of AW+SIC is depicted in Fig. \ref{AW_block_diagram}. Solid and dashed arrows denote data and control paths, respectively. The AW+SIC controller determines the decoding assignments based on the decoding results of BS~$i$ and BS~$j$ and feeds them into the decoding engine.  

Let us denote the SINR at the BSs corresponding to anywhere decoding by $\text{SINR}^{\rm{AW+SIC}}$. Using AW+SIC, we will have four possibilities for the SINR of UE~1, i.e., $\text{SINR}_1^{\rm{AW+SIC}}$, given~as

\begin{equation}\label{sinraw}
\text{SINR}_1^{\rm{AW+SIC}} =
\left\{
\begin{array}{ll}
\max\left\{P|h_{11}|^2,P|h_{12}|^2\right\}  & \mbox{if}\quad \mathcal{E}_{21} \cap \mathcal{E}_{22}\\
\max\left\{P|h_{11}|^2,\frac{P|h_{12}|^2}{1+P|h_{22}|^2}\right\}  & \mbox{if}\quad \mathcal{E}_{21} \cap \mathcal{E}_{22}^c\\
\max\left\{\frac{P|h_{11}|^2}{1+P|h_{21}|^2},P|h_{12}|^2\right\}  & \mbox{if}\quad \mathcal{E}_{21}^c \cap \mathcal{E}_{22}\\
\max\left\{\frac{P|h_{11}|^2}{1+P|h_{21}|^2},\frac{P|h_{12}|^2}{1+P|h_{22}|^2}\right\}  & \mbox{if}\quad \mathcal{E}_{21}^c \cap \mathcal{E}_{22}^c.\\
\end{array}
\right.
\end{equation}
\begin{table}[t]
\small
	\caption{AW+SIC ALGORITHM FOR TWO COOPERATING CELLS.} \label{awalgorithm}
	\centering
	\begin{tabular}{ |l|l|l| }
		\hline
		\multirow{2}{*}{Current Decoding Assignment} & Dec. & Decision/ Next\\
		& Result & Decoding Assignment\\ \hline
		\multicolumn{3}{|c|}{STEP 1} \\
		\hline
		\multirow{4}{*}{BS~$i$: UE~$i$/UE~$j$, BS~$j$: UE~$j$/UE~$i$} & 11 & Both UEs decoded; finish decoding\\
		& 10 & BS~$i$: UE~$j$, BS~$j$: UE~$i$/UE~$j$\\
		& 01 & BS~$i$: UE~$j$/UE~$i$, BS~$j$: UE~$i$\\
		& 00 & BS~$i$: UE~$j$/UE~$i$, BS~$j$: UE~$i$/UE~$j$\\ \hline
		\multicolumn{3}{|c|}{STEP 2} \\
		\hline
		\multirow{4}{*}{BS~$i$: UE~$j$, BS~$j$: UE~$i$/UE~$j$} & 11 & Both UEs decoded; finish decoding\\
		& 10 & Both UEs decoded; finish decoding\\
		& 01 & BS~$j$: UE~$j$\\
		& 00 & BS~$i$: Stop; UE~$j$ not decodable\\ \hline
		\multirow{4}{*}{BS~$i$: UE~$j$/UE~$i$, BS~$j$: UE~$i$} & 11 & Both UEs decoded; finish decoding\\
		& 10 & BS~$i$: UE~$i$\\
		& 01 & Both UEs decoded; finish decoding\\
		& 00 & BS~$i$: Stop; UE~$i$ not decodable\\ \hline
		\multirow{4}{*}{BS~$i$: UE~$j$/UE~$i$, BS~$j$: UE~$i$/UE~$j$} & 11 & Both UEs decoded; finish decoding\\
		& 10 & BS~$i$: UE~$i$\\
		& 01 & BS~$j$: UE~$j$\\
		& 00 & BS~$i$: Stop; both UEs not decodable\\ \hline
		\multicolumn{3}{|c|}{STEP 3} \\
		\hline
		\multirow{2}{*}{BS~$i$: UE~$i$} & 1 & Both UEs decoded; finish decoding\\
		& 0 & Stop; UE~$i$ not decodable\\ \hline
		\multirow{2}{*}{BS~$j$: UE~$j$} & 1 & Both UEs decoded; finish decoding\\
		& 0 & Stop; UE~$j$ not decodable\\ \hline
	\end{tabular}
\end{table}
To understand \eqref{sinraw}, let us assume that UE~2 is decodable at both BS~1 and BS~2 (treating the signal coming from UE~1 as noise). In this case, we will have two possibilities for the SINR of UE~1: $P|h_{11}|^2$ and $P|h_{12}|^2$, and therefore, the best SINR that we can achieve is $\max\left\{P|h_{11}|^2,P|h_{12}|^2\right\}$, if we allow for the signal of UE~1 to be decoded at any of the BSs. Additionally, if UE~2 is decodable at BS~1, but not BS~2 (treating the signal coming from UE~1 as noise), then there will be two possibilities for the SINR of UE~1: $P|h_{11}|^2$ at BS~1, by subtracting the signal coming from UE~2, and $\frac{P|h_{12}|^2}{1+P|h_{22}|^2}$ at BS~2, by treating the signal coming from UE~2 as noise. Hence, the best SINR that we can achieve in this case is $\max\left\{P|h_{11}|^2,\frac{P|h_{12}|^2}{1+P|h_{22}|^2}\right\}$. Similar arguments can be made for the third and fourth SINR expressions in \eqref{sinraw}. 
Based upon the discussion so far, the outage event for UE~1, using the AW+SIC scheme would be
\begin{equation}
\mathcal{O}_1^{\rm{AW+SIC}}=\mathcal{A}_{11}^c\cap \mathcal{A}_{12}^c.
\end{equation}
Now, we derive the outage probability for UE~1, using the AW+SIC scheme
\begin{align}\label{awout}
\mathbb{P}\left(\mathcal{O}_1^{\rm{AW+SIC}}\right)&=\mathbb{P}(\mathcal{A}_{11}^c)\mathbb{P}(\mathcal{A}_{12}^c),
\end{align}
which results from the fact that $A_{11}$ and $A_{12}$ are independent. After computing the probabilities of these events, the outage probability for UE~1 would be
\begin{align}\label{awout2}
P_{\rm{out}_1}^{\rm{AW+SIC}}(\theta_1)
&=\left\{
\begin{array}{ll}
f_{\rm{AW+SIC}}(\theta_1)  & \mbox{if}\quad \theta_1\geq1/\theta_2\\
g_{\rm{AW+SIC}}(\theta_1) & ~~\quad \text{otherwise},
\end{array}
\right.
\end{align}
where
\begin{align*}
f_{\rm{AW+SIC}}(\theta_1)&=f(\lambda_{11},\lambda_{21},\theta_2;\theta_1)f(\lambda_{12},\lambda_{22},\theta_2;\theta_1),\\
g_{\rm{AW+SIC}}(\theta_1)&=g(\lambda_{11},\lambda_{21},\theta_2;\theta_1)g(\lambda_{12},\lambda_{22},\theta_2;\theta_1).
\end{align*}
Asymptotically, the outage probability for UE~1 is
\begin{align}
P_{\rm{out}_1}^{\rm{AW+SIC}}(\theta_1)
&\overset{\theta_1\rightarrow 0}\sim\left(\frac{1-{\rm e}^{-\frac{\lambda_{21}\theta_2}{P}}}{\lambda_{21}}+\frac{1-\theta_2{\rm e}^{-\frac{\lambda_{21}\theta_2}{P}}}{P}\right)
\left(\frac{1-{\rm e}^{-\frac{\lambda_{22}\theta_2}{P}}}{\lambda_{22}}+\frac{1-\theta_2{\rm e}^{-\frac{\lambda_{22}\theta_2}{P}}}{P}\right)\lambda_{11}\lambda_{12}\theta_1^2.
\end{align}
For the symmetric case $\theta_1=\theta_2=\theta$, the asymptotic outage probability is
\begin{equation}\label{awasymp2}
P_{\rm{out}_1}^{\rm{AW+SIC}}(\theta)\overset{\theta\rightarrow 0}\sim\frac{\lambda_{11}\lambda_{12}}{P^2}\theta^2.
\end{equation}
\subsection{Anywhere Decoding and Distributed Interference Subtraction (AW+DIS)}\label{awdissec}
As we discussed in Section \ref{awdissec}, the conventional DIS scheme restricts UE~$i$ to be decoded at BS~$i, i\in\{1,2\}$. To further enhance the performance of conventional DIS, we can use anywhere decoding in combination with DIS: if the signal from a specific UE is decoded at \emph{any} of the BSs, the corresponding decoded message will be sent to the other BS through the backhaul. Let us denote the SINR at the BSs corresponding to anywhere decoding and DIS by $\text{SINR}^\text{AW+DIS}$. In this case, the SINR of UE~1, i.e., $\rm{SINR}_1^{\rm{AW+DIS}}$ is
\begin{equation}\label{4.17}
\rm{SINR}_1^{\rm{AW+DIS}} =
\left\{
\begin{array}{ll}
\max\left\{\frac{P|h_{11}|^2}{1+P|h_{21}|^2},\frac{P|h_{12}|^2}{1+P|h_{22}|^2}\right\}  & \mbox{if}\quad \mathcal{E}_{21}^c \cap \mathcal{E}_{22}^c\\
\max\left\{P|h_{11}|^2,P|h_{12}|^2\right\}  & ~~\quad \text{otherwise}.
\end{array}
\right.
\end{equation}
In words, if UE~2 is not decodable at any of the BSs (treating UE~1 as noise), we have two possibilities for the SINR of UE~1: $\frac{P|h_{11}|^2}{1+P|h_{21}|^2}$ and $\frac{P|h_{12}|^2}{1+P|h_{22}|^2}$, and through AW+DIS we get the maximum of the two values. Otherwise, if UE~2 is decodable by at least one of the BSs, UE~1 can be decoded at both BSs free from interference, as the decoded message gets exchanged between the BSs. 

\begin{figure}[t]
	\centering
	\includegraphics[width=5.2in]{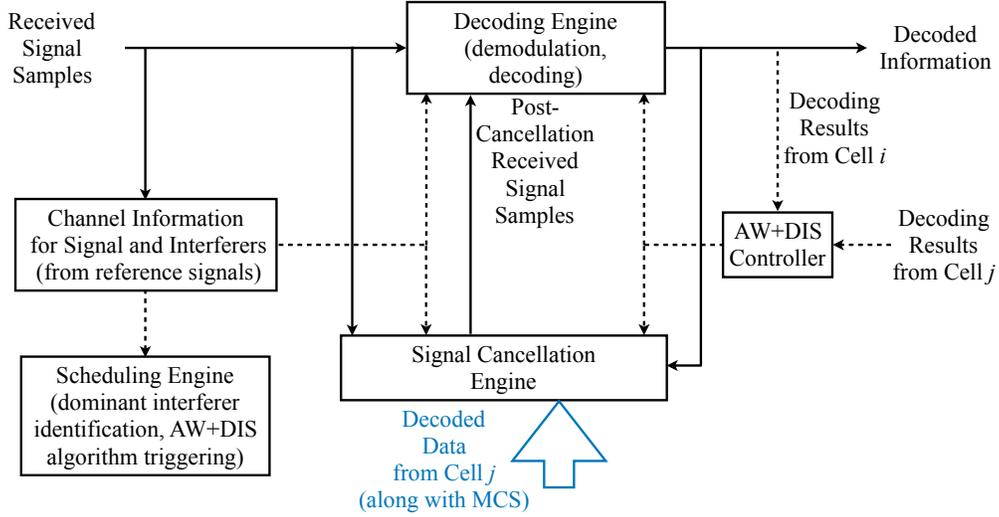}\\
	\vspace{-1cm}
	\caption{AW+DIS functional block diagram for BS~$i$. The AW+DIS controller determines the next decoding assignments based on the decoding results from the cooperating cells, and also determines which UE's data needs to be exchanged between the BSs.
	}\label{awdisbd}
\end{figure}

Now, we explain the AW+DIS algorithm in more detail, in the context of our 2-BS, 2-UE model. Let us again index the cooperating cells as $i$ and $j$. The three-step AW+DIS algorithm has been summarized in Table \ref{awdisalgorithm}, where a controller has an initial decoding assignment for BS~$i$ and BS~$j$, and determines the future decoding assignments based on the decoding results of previous steps. This controller could be a separate entity, connected to BS~$i$ and BS~$j$, or it could be a part of each of the BSs (as in Fig. \ref{awdisbd}). In addition to the notations introduced for Table~\ref{awalgorithm} in the AW+SIC algorithm, we use the following concise notation in Table \ref{awdisalgorithm}.
\begin{itemize}
	\item ${\rm BS}~i\xrightarrow{{\rm UE}~k} {\rm BS}~j$ means BS~$i$ sends the data of UE~$k$ to BS~$j$.
\end{itemize}

\begin{table}[t!]
\small
	\caption{AW+DIS ALGORITHM FOR TWO COOPERATING CELLS.} \label{awdisalgorithm}
	\centering
	\begin{tabular}{ |l|l|l| }
		\hline
		\multirow{2}{*}{Current Decoding Assignment} & Dec. & Decision/ Next\\
		& Result & Decoding Assignment\\ \hline
		\multicolumn{3}{|c|}{STEP 1} \\
		\hline
		\multirow{4}{*}{BS~$i$: UE~$i$/UE~$j$, BS~$j$: UE~$j$/UE~$i$} & 11 & Both UEs decoded; finish decoding\\
		& 10 & BS~$i$: UE~$j$, BS~$j$: UE~$i$/UE~$j$\\
		& 01 & BS~$i$: UE~$j$/UE~$i$, BS~$j$: UE~$i$\\
		& 00 & BS~$i$: UE~$j$/UE~$i$, BS~$j$: UE~$i$/UE~$j$\\ \hline
		\multicolumn{3}{|c|}{STEP 2} \\
		\hline
		\multirow{4}{*}{BS~$i$: UE~$j$, BS~$j$: UE~$i$/UE~$j$} & 11 & Both UEs decoded; finish decoding\\
		& 10 & Both UEs decoded; finish decoding\\
		& 01 & BS~$j$: UE~$j$\\
		& 00 & ${\rm BS}~i\xrightarrow{{\rm UE}~i} {\rm BS}~j$, BS~$j$: UE~$j$\\ \hline
		\multirow{4}{*}{BS~$i$: UE~$j$/UE~$i$, BS~$j$: UE~$i$} & 11 & Both UEs decoded; finish decoding\\
		& 10 & BS~$i$: UE~$i$\\
		& 01 & Both UEs decoded; finish decoding\\
		& 00 & ${\rm BS}~j\xrightarrow{{\rm UE}~j} {\rm BS}~i$, BS~$j$: UE~$j$\\ \hline
		\multirow{4}{*}{BS~$i$: UE~$j$/UE~$i$, BS~$j$: UE~$i$/UE~$j$} & 11 & Both UEs decoded; finish decoding\\
		& 10 & ${\rm BS}~i\xrightarrow{{\rm UE}~j} {\rm BS}~j$, BS~$i$: UE~$i$, BS~$j$: UE~$i$\\
		& 01 & ${\rm BS}~j\xrightarrow{{\rm UE}~i} {\rm BS}~i$, BS~$i$: UE~$j$, BS~$j$: UE~$j$\\
		& 00 & BS~$i$: Stop; both UEs not decodable\\ \hline
		\multicolumn{3}{|c|}{STEP 3} \\
		\hline
		\multirow{2}{*}{BS~$i$: UE~$i$, BS~$j$: UE~$i$} & 11,01,10 & Both UEs decoded; finish decoding\\
		& 00 & Stop; UE~$i$ not decodable\\ \hline
		\multirow{2}{*}{BS~$i$: UE~$j$, BS~$j$: UE~$j$} & 11,01,10 & Both UEs decoded; finish decoding\\
		& 00 & Stop; UE~$j$ not decodable\\ \hline
	\end{tabular}
\end{table}
A functional block diagram of AW+DIS is depicted in Fig. \ref{awdisbd}. Solid and dashed arrows demonstrate data and control paths, respectively. The AW+DIS controller determines the decoding assignments based on the decoding results of BS~$i$ and BS~$j$ and feeds them into the decoding engine. The distinction between AW+SIC and AW+DIS is that the decoded data can be exchanged between the BSs in the AW+DIS scheme, as depicted in Fig. \ref{awdisbd}.

Accordingly, the outage event for UE~1 using the AW+DIS scheme can be written as
\begin{equation}\nonumber
\mathcal{O}_1^{\rm{AW+DIS}}=\mathcal{A}_{11}^c\cap \mathcal{A}_{12}^c\cap \left(\mathcal{A}_{22}^c\cup \{P|h_{11}|^2<\theta_1\}\right).
\end{equation}

The outage probability for UE~1, using the AW+DIS scheme is then
\begin{align*}
\mathbb{P}(\mathcal{O}_1^{\rm{AW+DIS}})&=\mathbb{P}(\mathcal{A}_{11}^c\mathcal{A}_{12}^c\mathcal{A}_{22}^c)+\mathbb{P}(\mathcal{A}_{11}^c\mathcal{A}_{12}^c\cap \{P|h_{11}|^2<\theta_1\})-\mathbb{P}(\mathcal{A}_{11}^c\mathcal{A}_{12}^c\mathcal{A}_{22}^c\cap \{P|h_{11}|^2<\theta_1\})\\
&=\mathbb{P}(\mathcal{A}_{11}^c)\mathbb{P}(\mathcal{A}_{12}^c\mathcal{A}_{22}^c)+\mathbb{P}(\mathcal{A}_{12}^c)\mathbb{P}(P|h_{11}|^2<\theta_1)-\mathbb{P}(P|h_{11}|^2<\theta_1)\mathbb{P}(\mathcal{A}_{12}^c\mathcal{A}_{22}^c)\\
&=\left\{
\begin{array}{ll}
f_{\text{AW+DIS}}(\theta_1)  & \mbox{if}\quad \theta_1\geq1/\theta_2\\
g_{\text{AW+DIS}}(\theta_1) & ~~\quad \text{otherwise},
\end{array}
\right.
\end{align*}
where
\begin{align*}
f_{\rm{AW+DIS}}(\theta_1)&=f_{\rm{AW}}(\theta_1)+e(\lambda_{12},\lambda_{22},\theta_2;\theta_1)\left(1-{\rm e}^{-\frac{\lambda_{11}\theta_1}{P}}-f(\lambda_{12},\lambda_{22},\theta_2;\theta_1)\right),\\
g_{\rm{AW+DIS}}(\theta_1)&=g_{\rm{AW}}(\theta_1)+e(\lambda_{12},\lambda_{22},\theta_2;\theta_1)\left(1-{\rm e}^{-\frac{\lambda_{11}\theta_1}{P}}-g(\lambda_{12},\lambda_{22},\theta_2;\theta_1)\right),\\
e(\lambda_{12},\lambda_{22},\theta_2;\theta_1)&=\frac{\lambda_{12}{\rm e}^{-\frac{\lambda_{22}\theta_2}{P}}}{\lambda_{12}+\lambda_{22}\theta_2}\left(1-\exp\left(-\frac{(\lambda_{12}+\lambda_{22}\theta_2)\theta_1}{P}\right)\right).
\end{align*}
Now that we have the outage probability expressions for AW+SIC and AW+DIS schemes, in the following subsection, we compare their performance to the interference mitigation schemes discussed in Section \ref{noi}.
\subsection{Comparison of the Interference Mitigation Schemes}\label{comparison}
We consider the model depicted in Fig.~\ref{cellular}, and consider quasi-static Rayleigh fading and path loss. The cell radius and path loss exponent have been set to $d=2$, and $\alpha=4$, respectively. We assume that $\theta_1=\theta_2=\theta$,  and we compare the system performance for the six interference mitigation schemes described so far, for two scenarios. The outage probabilities for the decoding schemes have been plotted based on the analytical results for the MARP, AW+SIC, DIS, and AW+DIS schemes, and the plots for MIS and MMSE-SIC are based on simulations.

\begin{figure}[h]
	\centering
	\includegraphics[width=4.5in]{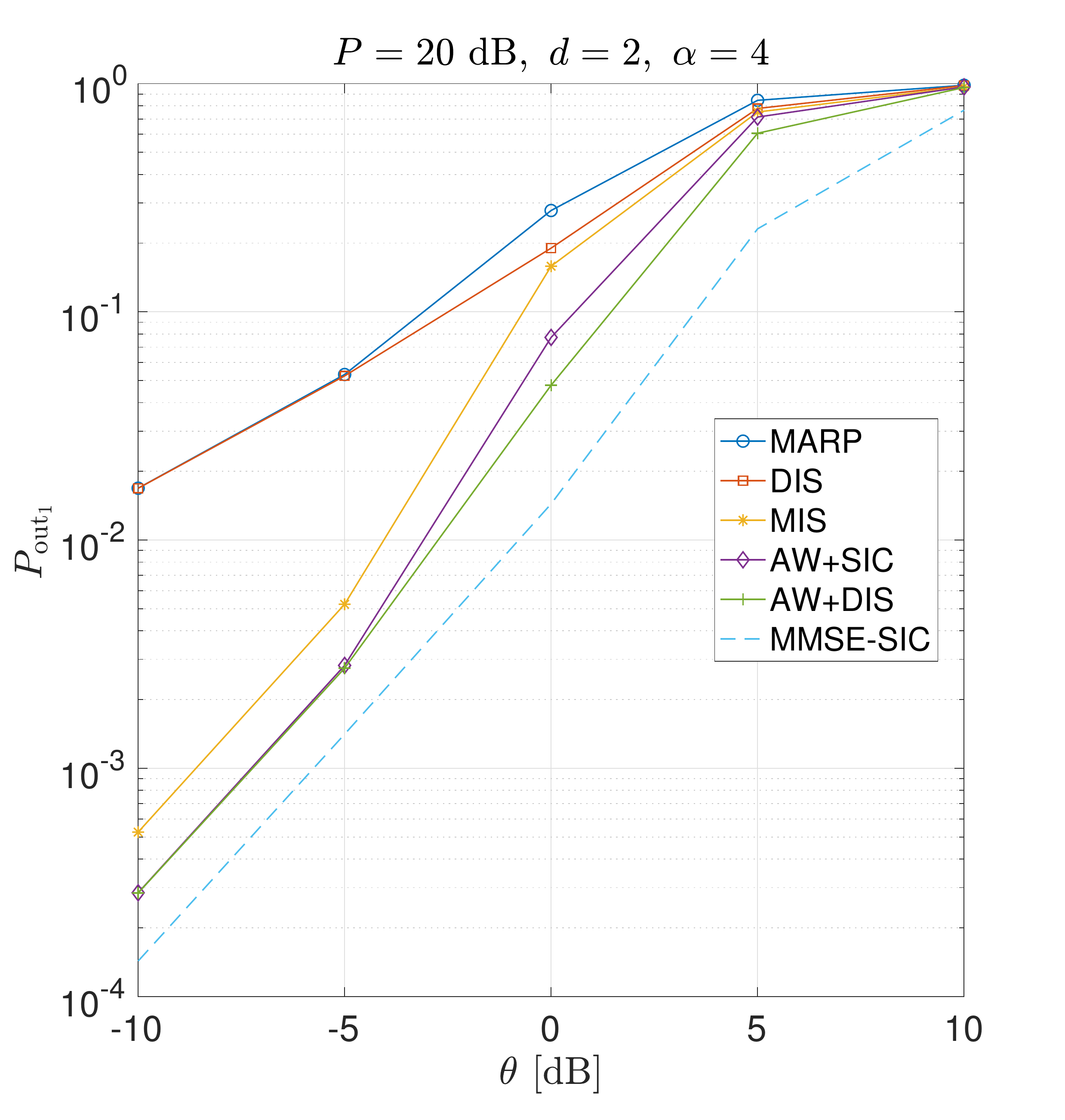}\\
	\caption{Outage probabilities of the six decoding schemes for UE~1, with both UEs located at cell edge, i.e., $z=d, t=-d$.
	}\label{celledge}
\end{figure}

As the first scenario, we assume that both of the UEs transmit with equal power, i.e. $P_1=P_2=P=20~\text{dB}$, and we consider an interference-limited scenario in which both of the UEs are located at the cell edge, i.e., $z=d, t=-d$. We refer to these UEs as \emph{worst-case UEs}. We can readily see from Fig.~\ref{celledge} that, for $R_1=R_2=1$ bit/sec/Hz ($\theta=0~\rm dB$), we observe a 72\% reduction in the outage probability for UE~1 if we use anywhere decoding instead of MARP. By using a combination of anywhere decoding and DIS, we obtain an additional 11\% reduction (83\% compared to MARP) in the outage probability. As $\theta\rightarrow 0$, we observe that there is approximately a 1.5 dB gap between the performances of the AW+SIC and MMSE-SIC schemes. The significance of this observation is that we can achieve a performance close to MMSE-SIC in the asymptotic regime while exchanging only a few bits among the BSs.

\begin{figure}[h]
	\centering
	\includegraphics[width=4.5in]{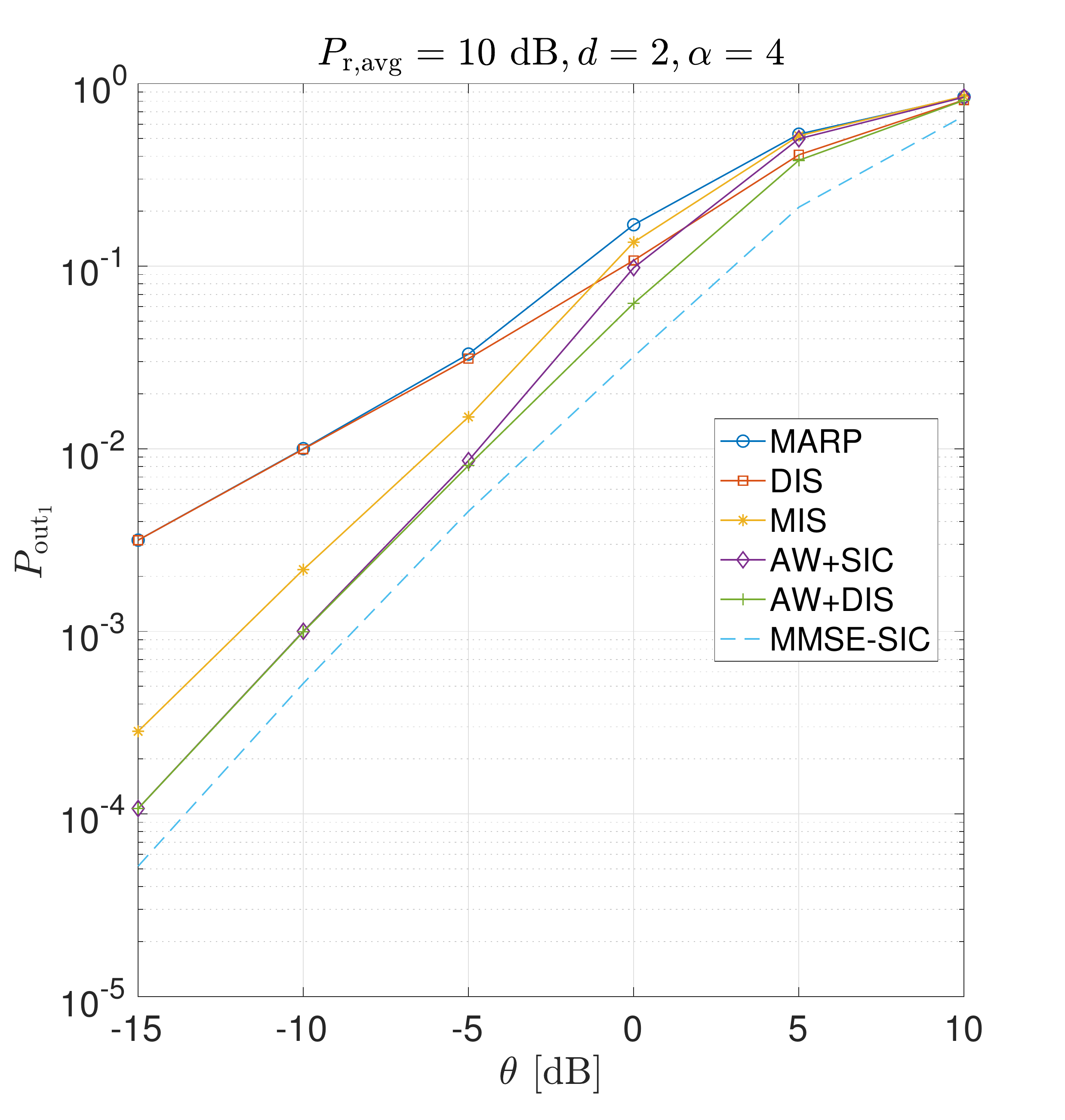}\\
	\caption{Outage probabilities of the six decoding schemes for UE~1, with UEs located randomly and uniformly from halfway to their respective base stations to the common cell edge, i.e., $Z\sim \mathcal{U}[d/2,d]$ for UE~1, $T\sim \mathcal{U}[-d,-d/2]$ for UE~2.
	}\label{powercontrol_half}
\end{figure}

For the second scenario, we assume random UE locations, from halfway from their respective base stations to the common cell edge, i.e., $Z\sim \mathcal{U}[d/2,d]$ for UE~1 and $T\sim \mathcal{U}[-d,-d/2]$ for UE~2. We further assume that the UEs perform power control for the purpose of full path loss compensation, so that the average received powers at their associated BSs are $P_{\mathrm{r,avg}}=10~\text{dB}$. Outage probabilities of the six decoding schemes for UE~1 are plotted in Fig.~\ref{powercontrol_half}. For the symmetric case in which $R_1=R_2=1$ bit/sec/Hz ($\theta=0~\rm dB$), the AW+SIC and AW+DIS schemes lead to 42\% and 63\% reductions in the outage probability, respectively. As $\theta \rightarrow 0$, we again observe that there is approximately a 1.5 dB gap between the performance of AW+SIC and MMSE-SIC schemes, which suggests that this gap is independent of the respective positioning of the UEs as well as their transmit power. Generally, we observe more pronounced gains for AW+SIC and AW+DIS schemes if the UEs are located closer to the cell edge.


\section{Performance of Uplink Interference Mitigation Schemes in the presence of Interferers Outside the Cooperating Cells}\label{ppp}
Similar to Section \ref{noi}, we first perform an outage probability analysis for the case in which the locations of the users are fixed and there is no power control, in Section \ref{npcfli}. Later,  in Section \ref{pcrli}, we generalize the results to incorporate randomness in user locations as well as power control. We primarily focus on deriving outage probability for the MARP scheme, which is the baseline scheme, and also the anywhere decoding (AW+SIC) scheme.
\subsection{Fixed UE Locations, No Power Control}\label{npcfli}
Let us consider the system model depicted in Fig.~\ref{cellular}, and also assume a one-dimensional PPP of interferers with intensity $\lambda$ in the interval $\mathbb{R}\setminus(-2d,2d)$, i.e., outside the two cooperating cells. We assume independent Rayleigh fading and that all UEs use transmit power $P$. In this subsection, we assume fixed locations for the two UEs in the cooperating cells.

The outage expressions derived in Sections \ref{noi} and \ref{awdecoding} can be reused by conditioning on the interference coming from the Poisson field of interferers and then averaging over the interference. Let $I_1$ be the interference at BS~1 coming from Poisson interferers if they have unit transmit power. In this case, the SINR for UE~1, treating UE~2 and the interference coming from the Poisson field of interferers as noise, is
\begin{align}
\mathrm{SINR}_1=\frac{P|h_{11}|^2}{1+PI_1+P|h_{21}|^2}=\frac{\frac{P}{1+PI_1}|h_{11}|^2}{1+\frac{P}{1+PI_1}|h_{21}|^2}.
\end{align}
In other words, we can obtain the outage expressions in the presence of a Poisson field of interferers by substituting $\frac{P}{1+PI_1}$ for $P$, and then averaging over $I_1$. With this being said, we analyze the outage for MARP and AW+SIC schemes in the following two subsections.
\paragraph{Outage Analysis for MARP}
Following the notations in Section \ref{sysmod}, $g_{x,1}$ denotes the Rayleigh fading complex channel gain from point $x$ to BS~1, $G_{x,1}=|g_{x,1}|^2$ ($\mathbb{E}[G_{x,1}]=1$), and we use the path loss model introduced therein. We use the outage expression in \eqref{marpout2}, and average over $I_1$, but before doing so, we define several functions to make the outage expressions more concise. Let
\begin{align*}
	\mathcal{L}_1(s)&\triangleq\mathbb{E}\left[e^{-sI_1}\right]=\mathbb{E}\exp\left(-\sum\limits_{x\in\Phi}\frac{sG_{x,1}}{1+|x+d|^\alpha}\right)\overset{\rm (a)}=\mathbb{E}\prod\limits_{x\in\Phi}\mathbb{E}_G\exp\left(-\frac{sG_{x,1}}{1+|x+d|^\alpha}\right)\\
	&=\mathbb{E}\prod\limits_{x\in\Phi}\frac{1}{1+\frac{s}{1+|x+d|^\alpha}}\overset{\rm (b)}=\exp\left(-\lambda\int_{\mathbb{R}\setminus(-2d,2d)}\left(1-\frac{1+|x+d|^{\alpha}}{s+1+|x+d|^{\alpha}}\right){\rm d}x\right),\\
	\mathcal{L}_2(s)
	&=\exp\left(-\lambda\int_{\mathbb{R}\setminus(-2d,2d)}\left(1-\frac{1+|x-d|^{\alpha}}{s+1+|x-d|^{\alpha}}\right){\rm d}x\right),
\end{align*}	
where $\mathcal{L}_1(s)$ and $\mathcal{L}_2(s)$ are Laplace transforms of $I_1$ and $I_2$, respectively. Here (a) follows from the independence of the fading random variables $G_{x,1}$, and (b) follows from the probability generating functional of the PPP \cite{haenggi_2012}. Furthermore, define
\begin{align}
K(a,b,\theta_1,\theta_2)\triangleq a\theta_1+b\theta_2(1+\theta_1), \quad
W(a,b,\theta_1,\theta_2)\triangleq\frac{b\theta_2(1+\theta_1)+a\theta_1(1+\theta_2)}{1-\theta_1\theta_2}.\nonumber
\end{align}	
The outage probability for the MARP scheme is then
\begin{align}\label{icoutage}
P_{\text{out}_1}^{\rm{MARP}}(\theta_1)&=\left\{
\begin{array}{ll}
f_I^{\rm MARP}(\lambda_{11},\lambda_{21},P,\theta_2;\theta_1)  & \mbox{if}\quad \theta_1\geq 1/\theta_2\\
g_I^{\rm MARP}(\lambda_{11},\lambda_{21},P,\theta_2;\theta_1) & ~~\quad \text{otherwise},
\end{array}
\right.
\end{align}
where
\begin{align*}
f_I^{\rm MARP}(a,b,P,\theta_2;\theta_1)&=\mathbb{E}_{I_1}\left[f\left(a,b,\frac{P}{1+PI_1},\theta_2;\theta_1\right)\right]
=1-\frac{b{\rm e}^{-\frac{a\theta_1}{P}}\mathcal{L}_1(a\theta_1)}{a\theta_1+b}-\frac{a{\rm e}^{-\frac{K}{P}}\mathcal{L}_1\left(K\right)}{a+b\theta_2},\\
g_I^{\rm MARP}(a,b,P,\theta_2;\theta_1)&=\mathbb{E}_{I_1}\left[g\left(a,b,\frac{P}{1+PI_1},\theta_2;\theta_1\right)\right]\\
&=f_I^{\rm MARP}(a,b,P,\theta_2;\theta_1)+\frac{ab(1-\theta_1\theta_2){\rm e}^{-\frac{W}{P}}\mathcal{L}_1(W)}{(a+b\theta_2)(a\theta_1+b)}.
\end{align*}
Using \eqref{awasymp2}, the asymptotic outage probability for the symmetric case, i.e., $\theta_1=\theta_2=\theta$, is
\begin{align}\label{marpasympi}
P_{\rm{out}_1}^{\rm{MARP}}(\theta)\overset{\theta \rightarrow 0}\sim\mathbb{E}_{I_1}\left[\lambda_{11}\left(I_1+\frac{1}{P}\right)\theta\right]
=\frac{\lambda_{11}}{P}\left(P\mathbb{E}[I_1]+1\right)\theta,
\end{align}
where
\begin{align}\label{meani}
\mathbb{E}[I_1]=\mathbb{E}\sum\limits_{x\in\Phi}\left(\frac{G_{x,1}}{1+|x+d|^\alpha}\right)
=\mathbb{E}\sum\limits_{x\in\Phi}\left(\frac{\mathbb{E}_G[G_{x,1}]}{1+|x+d|^\alpha}\right)
\overset{\rm (a)}=\lambda\int_{\mathbb{R}\setminus(-2r,2r)}\frac{{\rm d}x}{1+|x+d|^\alpha},
\end{align}
and (a) follows from Campbell's theorem for sums \cite{haenggi_2012}. Comparing \eqref{marpasympi} to \eqref{awasymp2}, we observe that there is an increase in the outage probability by a factor of $P\mathbb{E}[I_1]+1$, and we note that this factor is independent of the respective positioning of the UEs. 
\paragraph{Outage Analysis for Anywhere Decoding with SIC}\label{awpppout}
We define some additional functions to make the outage expressions more concise. Let
\begin{align*}
L(c,d,\theta_1,\theta_2)\triangleq c\theta_1+d\theta_2(1+\theta_1),\quad
V(c,d,\theta_1,\theta_2)\triangleq\frac{d\theta_2(1+\theta_1)+c\theta_1(1+\theta_2)}{1-\theta_1\theta_2},\nonumber
\end{align*}
\begin{align*}
\mathcal{L}(s_1,s_2)
&\triangleq\mathbb{E}\left[{\rm e}^{-s_1I_1}{\rm e}^{-s_2I_2}\right]\nonumber
=\mathbb{E}\exp\left(-s_1\sum\limits_{x\in\Phi}\left(\frac{G_{x,1}}{1+|x+d|^\alpha}\right)-s_2\sum\limits_{x\in\Phi}\left(\frac{G_{x,2}}{1+|x-d|^\alpha}\right)\right)\nonumber\\
&=\mathbb{E}\prod\limits_{x\in\Phi}\mathbb{E}_G\exp\left(-s_1\left(\frac{G_{x,1}}{1+|x+d|^\alpha}\right)-s_2\left(\frac{G_{x,2}}{1+|x-d|^\alpha}\right)\right)\nonumber\\
&=\mathbb{E}\prod\limits_{x\in\Phi}\frac{1}{\left(1+\frac{s_1}{1+|x+d|^\alpha}\right)\left(1+\frac{s_2}{1+|x+d|^\alpha}\right)}\nonumber\\
&=\exp\left(-\lambda\int_{\mathbb{R}\setminus(-2d,2d)}\left(1-\frac{(1+|x+d|^{\alpha})(1+|x-d|^{\alpha})}{\left(s_1+1+|x+d|^{\alpha}\right)\left(s_2+1+|x-d|^{\alpha}\right)}\right){\rm d}x\right),
\end{align*}
where $\mathcal{L}(s_1,s_2)$ is the joint Laplace transform of $I_1$ and $I_2$. With these defined, the outage probability for AW+SIC can be derived in a similar way to MARP, leading to
\begin{align}
P_{\text{out}_1}^{\text{AW+SIC}}(\theta_1)&=\left\{
\begin{array}{ll}\label{awoutage}
f_I^{\rm AW+SIC}(\lambda_{11},\lambda_{21},\lambda_{12},\lambda_{22},\theta_2;\theta_1)  & \mbox{if}\quad \theta_1\geq 1/\theta_2\\
g_I^{\rm AW+SIC}(\lambda_{11},\lambda_{21},\lambda_{12},\lambda_{22},\theta_2;\theta_1) & ~~\quad \text{otherwise},
\end{array}
\right.
\end{align}
where\footnote{$f_I^{\rm AW+SIC}(a,b,c,d,\theta_2;\theta_1)$ and $g_I^{\rm AW+SIC}(a,b,c,d,\theta_2;\theta_1)$ have been abbreviated as $f_I^{\rm AW+SIC}(\theta_1)$ and $g_I^{\rm AW+SIC}(\theta_1)$, respectively.}
\begin{align*}
f_I^{\rm AW+SIC}(\theta_1)=&~ \mathbb{E}_{I_1I_2}\left[f\left(a,b,\frac{P}{1+PI_1},\theta_2;\theta_1\right)f\left(c,d,\frac{P}{1+PI_2},\theta_2;\theta_1\right)\right]\\
=&~f_I^{\rm MARP}(a,b,\theta_2;\theta_1)-\frac{d{\rm e}^{-\frac{c\theta_1}{P}}\mathcal{L}_2(c\theta_1)}{c\theta_1+d}-\frac{c{\rm e}^{-\frac{L}{P}}\mathcal{L}_1(L)}{c+d\theta_2}
+\frac{bd{\rm e}^{-\frac{(a+c)\theta_1}{P}}\mathcal{L}(a\theta_1,c\theta_1)}{(a\theta_1+b)(c\theta_1+d)}\\
&+\frac{bc{\rm e}^{-\frac{a\theta_1+L}{P}}\mathcal{L}(a\theta_1,L)}{(a\theta_1+b)(c+d\theta_2)}
+\frac{ad{\rm e}^{-\frac{K+c\theta_1}{P}}\mathcal{L}(K,c\theta_1)}{(a+b\theta_2)(c\theta_1+d)}+\frac{ac{\rm e}^{-\frac{K+L}{P}}\mathcal{L}(K,L)}{(a+b\theta_2)(c+d\theta_2)},
\end{align*}
\begin{align*}
g_I^{\rm AW+SIC}(\theta_1)=&~ \mathbb{E}_{I_1I_2}\left[g\left(a,b,\frac{P}{1+PI_1},\theta_2;\theta_1)g(c,d,\frac{P}{1+PI_2},\theta_2;\theta_1\right)\right]=f_I^{\rm AW}(a,b,c,d,P,\theta_2;\theta_1)\\
&+\frac{cd(1-\theta_1\theta_2){\rm{e}}^{-\frac{V}{P}}}{(c+d\theta_2)(c\theta_1+d)}\times\left(\mathcal{L}_2(V)-\frac{b{\rm{e}}^{-\frac{a\theta_1}{P}}\mathcal{L}(a\theta_1,V)}{a\theta_1+b}-\frac{a{\rm{e}}^{-\frac{K}{P}}\mathcal{L}(K,V)}{a+b\theta_2}\right)\\
&+\frac{ab(1-\theta_1\theta_2){\rm{e}}^{-\frac{W}{P}}}{(a+b\theta_2)(a\theta_1+b)}\times\left(\mathcal{L}_1(W)-\frac{d{\rm{e}}^{-\frac{c\theta_1}{P}}\mathcal{L}(W,c\theta_1)}{c\theta_1+d}-\frac{c{\rm{e}}^{-\frac{L}{P}}\mathcal{L}(W,L)}{c+d\theta_2}\right)\\
&+\frac{abcd(1-\theta_1\theta_2)^2}{(a+b\theta_2)(a\theta_1+b)(c+d\theta_2)(c\theta_1+d)}{\rm{e}}^{-\frac{V+W}{P}}\mathcal{L}(W,V).
\end{align*}
Using \eqref{awasymp2}, the asymptotic outage probability for the symmetric case $\theta_1=\theta_2=\theta$ becomes
\begin{align*}
P_{\rm{out}_1}^{\rm{AW+SIC}}(\theta)&\overset{\theta\rightarrow 0}\sim\mathbb{E}\left[\lambda_{11}\lambda_{12}\left(I_1+\frac{1}{P}\right)\left(I_2+\frac{1}{P}\right)\theta^2\right]=\frac{\lambda_{11}\lambda_{12}}{P^2}\left(P^2\mathbb{E}[I_1I_2]+P(\mathbb{E}[I_1]+\mathbb{E}[I_2])+1\right)\theta^2,
\end{align*}
where $\mathbb{E}[I_1]$ is derived in \eqref{meani}, 
\begin{align*}\label{ei1i2}
\mathbb{E}[I_2]&=\lambda\int_{\mathbb{R}\setminus(-2d,2d)}\frac{{\rm d}x}{1+|x-d|^\alpha},
\end{align*}
and
\begin{align*}
\mathbb{E}[I_1I_2]=&~\mathbb{E}\left[\sum\limits_{x\in\Phi}\left(\frac{G_{x,1}}{1+|x+d|^\alpha}\right)\sum\limits_{x\in\Phi}\left(\frac{G_{x,2}}{1+|x-d|^\alpha}\right)\right]\\
\overset{\rm (a)}=&~\mathbb{E}\left[\sum\limits_{x\in\Phi}\left(\frac{\mathbb{E}_G[G_{x,1}]}{1+|x+d|^\alpha}\right)\sum\limits_{x\in\Phi}\left(\frac{\mathbb{E}_G[G_{x,2}]}{1+|x-d|^\alpha}\right)\right]\\
=&~\mathbb{E}\left[\sum\limits_{i}\left(\frac{1}{1+|x_i+d|^\alpha}\right)\sum\limits_{j}\left(\frac{1}{1+|x_j-d|^\alpha}\right)\right]\\
=&~\mathbb{E}\left[\sum\limits_{i}\left(\frac{1}{1+|x_i+d|^\alpha}\frac{1}{1+|x_i-d|^\alpha}+\frac{1}{1+|x_i+d|^\alpha}\sum\limits_{j\neq i}\frac{1}{1+|x_j-d|^\alpha}\right)\right]\\
\overset{\rm (b)}=&~\lambda\int_{\mathbb{R}\setminus(-2d,2d)}\frac{{\rm d}x}{(1+|x+d|^\alpha)(1+|x-d|^\alpha)}\\
&+\lambda^2\int_{\mathbb{R}\setminus(-2d,2d)}\frac{{\rm d}x}{1+|x+d|^\alpha}\int_{\mathbb{R}\setminus(-2d,2d)}\frac{{\rm d}x}{1+|x-d|^\alpha}\\
=&~\lambda\int_{\mathbb{R}\setminus(-2d,2d)}\frac{{\rm d}x}{(1+|x+d|^\alpha)(1+|x-d|^\alpha)}+\mathbb{E}[I_1]\mathbb{E}[I_2],
\end{align*}
and where (a) follows from the independence of $G_{x,1}$ and $G_{x,2}$, and (b) follows from Campbell's theorem for sums.

If the system is symmetric in terms of the out-of-cooperating-cell interferers, we have $\mathbb{E}[I_1]=\mathbb{E}[I_2]$. Additionally, if $\mathbb{E}[I_1I_2]\simeq \mathbb{E}[I_1]\mathbb{E}[I_2]$, which happens if there is limited spatial correlation between $I_1$ and $I_2$, we have 
\begin{equation}\label{awasympi2}
P_{\rm{out}_1}^{\rm{AW+SIC}}(\theta)
\sim\frac{\lambda_{11}\lambda_{12}}{P^2}\left(P\mathbb{E}[I_1]+1\right)^2\theta^2,
\end{equation}
which means that we should anticipate the same horizontal shifts, i.e., $10\log_{10}\left(P\mathbb{E}[I_1]+1\right)$ in the outage plots of both the baseline and AW+SIC schemes, whenever we have a PPP field of interferers outside the cooperating cells. We can infer this from \eqref{awasympi2}, \eqref{marpasympi}, \eqref{awasymp2}, and \eqref{marpasymp2}. 
\subsection{Random UE Locations with Power Control}\label{pcrli}
We consider the same system model as discussed in Section \ref{npcfli}, except for these two additions:
\begin{itemize}
	\item We consider uplink power control (full path loss compensation) for the users within the cooperating cells, i.e., cells 1 and 2, and we additionally assume that the PPP field of interferers with intensity $\lambda$ in the interval $R-(-2d,2d)$ transmit with maximum power. By \emph{maximum power}, we mean the power they would send if they performed power control and were located at their corresponding cell edges.  
	\item The two users under consideration, i.e., UE~1 and UE~2, are located uniformly at random in the interval from their cell centers to their cell edges. 
\end{itemize}

For this scenario, we can reuse the outage probability expressions derived in Section \ref{npcfli} by considering an equivalent system model. Instead of considering a transmit power of $P\lambda_{11}$ at UE~1 (which means its corresponding received power at BS~1 and BS~2 are $P$ and $P\lambda_{11}/\lambda_{12}$, respectively), we assume that we have an equivalent system in which the transmit power is always $P$, but the values for the corresponding path losses have been updated. Let us denote the updated values for $\lambda_{ij}$ with $\lambda_{ij}^P$, so that $\lambda_{11}^P=1$, $\lambda_{12}^P=\lambda_{12}/\lambda_{11}$, $\lambda_{21}^P=\lambda_{21}/\lambda_{22}$, and $\lambda_{22}^P=1$. Similarly, for the signals coming from the PPP field of interferers, we can assume that the transmit power is equal to $P$, and the corresponding path loss values are multiplied by $1+d^\alpha$. The reason for this approach is that we can reuse the expressions derived in the previous subsection, i.e., \eqref{icoutage} and \eqref{awoutage}, with fixed transmit power $P$, and incorporate the effect of power control in the updated path loss values. This is how we can incorporate power control in the outage probability derivations. To incorporate randomness in the locations of the UEs in the cooperating cells, we average over the locations of the UEs, as in Section \ref{rlpc}.
\vspace{-2mm}
\subsection{Outage Performance Comparison of AW+SIC and MARP Schemes}
We consider the model depicted in Fig.~\ref{cellular} and quasi-static Rayleigh fading and path loss. The cell length and path loss exponent have been set to $d=2$, and $\alpha=4$, respectively. Additionally, we assume a one-dimensional PPP of interferers with intensity $\lambda=0.25$ (one user per cell, on average) in $\mathbb{R}\setminus(-2d,2d)$. We assume that $\theta_1=\theta_2=\theta$,  and we compare the system performance for the MARP and AW+SIC schemes, considering two different scenarios.

\begin{figure}[t!]
	\centering
	\includegraphics[width=4.5in]{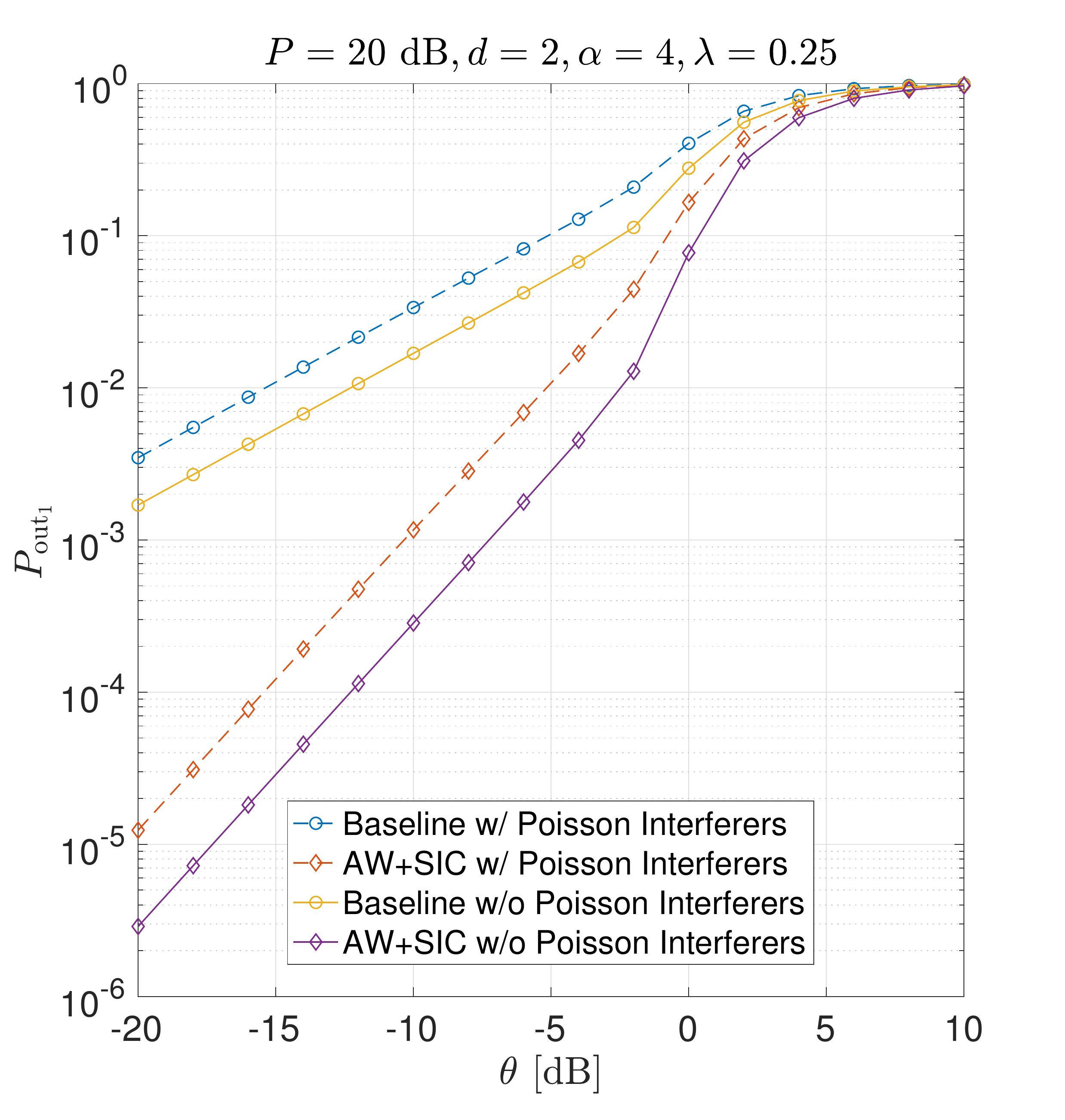}\\
	\caption{Outage probability of MARP and AW+SIC schemes w/ and w/o PPP interferers outside the cooperating cells, both UEs located at cell edge.
	}\label{fixeduserpe}
\end{figure}

First, we assume that both UEs (as well as the interferers outside the cooperating cells) transmit with equal power, i.e. $P_1=P_2=P=20~\text{dB}$, and we consider an interference-limited scenario, in which both of the UEs are located at the cell edge, i.e., $z=d, t=-d$. Fig.~\ref{fixeduserpe} illustrates the outage probability of the MARP and AW+SIC decoding schemes in the presence and absence of PPP interferers outside the cooperating cells. The plot in Fig.~\ref{fixeduserpe} is based on \eqref{icoutage} and \eqref{awoutage} for Poisson interferers and \eqref{marpoutsym} and \eqref{awout2} for no Poisson interferers outside the cooperating cells, respectively. If there is a PPP field of interferers outside the cooperating cells, there is a 59\% reduction in the outage probability if we use AW+SIC instead of MARP. On the other hand, as mentioned earlier in Section \ref{npcfli}, for the asymptotic regime, we have almost the same horizontal shift for both the MARP and AW+SIC schemes if we have PPP field of interferers, which is quantified by $10\log_{10}(P\mathbb{E}[I_1]+1)$.

\begin{figure}[t!]
	\centering
	\includegraphics[width=4.5in]{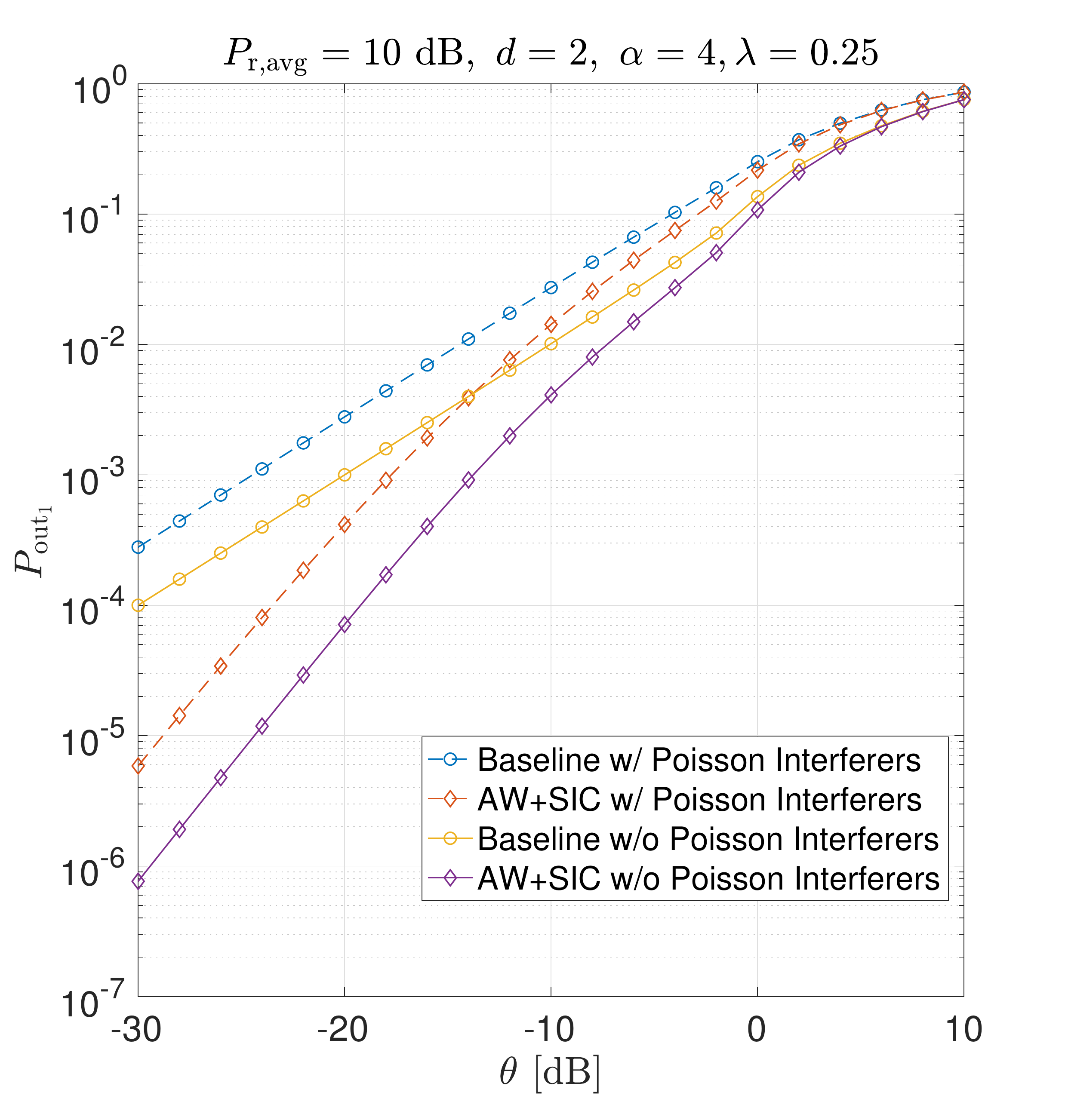}\\
	\caption{Outage probability of MARP and AW+SIC schemes w/ and w/o PPP interferers outside the cooperating cells, both UEs located uniformly at random from their corresponding cell centers to cell edges, i.e., $Z\sim \mathcal{U}[0,d], 
		T\sim \mathcal{U}[-d,0]$.
	}\label{randuseri}
\end{figure}

Second, we assume that the UEs in the cooperating cells are located uniformly at random in the interval from their cell centers to cell edges and perform power control. The target received power at the associated BSs for all cells is $10~\mathrm{dB}$. The outage probability for AW+SIC and MARP schemes have been depicted in Fig.~\ref{randuseri} in the presence and absence of the PPP field of interferers outside the cooperating cells. We observe the same horizontal shift in the asymptotic regime, both for the MARP and AW schemes, and also compared to Fig. \ref{fixeduserpe}, which is an indication of the fact that the value of the horizontal shift is independent of the respective positioning of the UEs within the cooperating cells, and also of the power control scheme utilized by the UEs.
\section{conclusion}
In this paper, a novel low-overhead uplink interference mitigation scheme has been explored for cellular systems. This scheme is based on the insight that for uplink transmissions it is not important at which BS the signal from a specific UE is decoded. We can leverage this fact by having flexible decoding assignments in which the cooperating BSs decode UEs collaboratively. We have shown considerable reductions in the outage probability relative to the baseline scheme with no BS cooperation, specifically for cell-edge UEs. Asymptotic results indicate that there is a 1.5 dB gap between the performance of anywhere decoding and full BS cooperation.

\bibliographystyle{IEEEtran}
\bibliography{AW_Journal}
\end{document}